\documentclass[aps,prl,twocolumn,groupedaddress,showpacs]{revtex4}

\usepackage{graphicx}
\usepackage{amsmath}

\begin{document}


\title{Dynamical vortex phases in a Bose-Einstein condensate driven by a rotating optical lattice}


\author{Kenichi Kasamatsu$^{1}$}
\author{Makoto Tsubota$^{2}$}

\affiliation{$^1$Department of General Education, 
Ishikawa National College of Technology, Tsubata, Ishikawa 929-0392, Japan\\
$^2$Department of Physics,
Osaka City University, Sumiyoshi-Ku, Osaka 558-8585, Japan}


\date{\today}

\begin{abstract}
We present simulation results of the vortex dynamics in a trapped Bose-Einstein condensate in the presence of a rotating optical lattice. Changing the potential amplitude and the relative rotation frequency between the condensate and the optical lattice, we find a rich variety of dynamical phases of vortices. The onset of these different phases is described by the force balance of a driving force, a pinning force and vortex-vortex interactions. In particular, when the optical lattice rotates faster than the condensate, an incommensurate effect leads to a vortex liquid phase supported by the competition between the driving force and the dissipation. 
\end{abstract}


\maketitle

Driven vortex lattices interacting with periodic pinning sites have attracted considerable interests, because they exhibit novel phase transitions between dynamical vortex states as a function of driving force. This problem has been studied in a wide variety of condensed matter systems including type-II superconductors and Josephson-junction arrays, described by the Frenkel-Kontrova-type models of friction \cite{Braun}. Recently, alkali-atomic Bose-Einstein condensates (BECs) in periodic potentials have been proposed to be another possible, superior system to study the vortex state caused by the interplay between pinning effects and vortex-vortex interactions \cite{Reijnders,Pu,Wu,Bhat,Tung}. In the condensed matter systems such as superconductors, the pinning effects incidental to unavoidable impurities play an essential role in the vortex dynamics. Since the BEC system is free from impurities, the intrinsic phenomena associated with quantized vortices can be studied through direct visualization \cite{Madison}. In addition, a laser-induced optical lattice (OL) constructs an ideal periodic potential, in which the microscopic pinning parameters, such as depth, periodicity, and symmetry of the lattice can be carefully controlled \cite{Morsch}. 

In this Letter, we study the dynamic properties of vortices in a trapped BEC driven by a rotating OL. A rotating OL has already been realized experimentally by using a laser beam passing through a rotating mask \cite{Tung}. In the previous studies \cite{Reijnders,Pu,Wu,Bhat}, the authors considered the equilibrium properties of a rotating BEC in the presence of a corotating OL with the {\it same} rotation frequency as the BEC. They obtained the phase diagram of the vortex lattice structure as a function of the pinning parameters through variational \cite{Reijnders} or numerical \cite{Pu} analysis. For a strongly interacting limit, this system can be mapped into the Bose-Hubbard Hamiltonian in the rotating frame \cite{Wu,Bhat}, exhibiting a series of quantum phase transition \cite{Bhat}. Tung {\it et al.} reported the experimental demonstration of the vortex pinning and the evidence of the structural phase transition \cite{Tung}. Beside these interesting studies, we can also consider a problem in which the rotation frequency of the OL is {\it different} from that of the condensate. Then, the moving optical lattice yields a rich variety of nonequilibrium vortex phases that are characterized by distinct vortex dynamics. The onset of these phases can be understood by the simple argument of a balance between the driving force, the pinning force, and the force coming from vortex-vortex interactions. In particular, when the optical lattice rotates faster than the condensate, the competition between the driving force and the dissipation causes a vortex liquid-like phase. This is an example of dissipative structures sustained by energy input and output \cite{Nicolis}, which offers an avenue to study them in atomic-gas BECs.

This system allows us to run {\it ab-initio} simulations based on the Gross-Pitaevskii (GP) equation \cite{Pethickbook}, whereas the other system has relied on phenomenological theory \cite{Braun}. Suppose that the condensate is created in a rotating harmonic trap with an angular velocity $\Omega \hat{\bf z}$. Here, we confine ourselves to the pancake geometry by assuming a tight axial confinement of a harmonic trap and high rotation rates which reduce the radial trap frequency $\omega_{\perp}$. We then give the Gaussian for the frozen axial component of the condensate wave function, obtaining a quasi-two-dimensional GP equation with a renormalized coupling constant. In a frame rotating with $\Omega$, the transverse wave function $\psi(x,y,t)$ (normalized to be unity) obeys 
\begin{equation}
(i - \gamma) \frac{\partial \psi}{\partial t} = \left[ -\nabla^{2} + \frac{r^{2}}{4} + V_{\rm OL} + U |\psi|^{2} - \mu - \Omega L_{z} \right] \psi; \label{tdGPeq}
\end{equation}
we choose our units for time, length, and energy as $1/\omega_{\perp}$, $a_{\perp}=\sqrt{\hbar/2 m\omega_{\perp}}$, and $\hbar \omega_{\perp}$, respectively. Here, $r^{2} = x^{2} + y^{2}$, $m$ the atomic mass, $\mu$ the chemical potential, and $L_{z}$ the $z$ component of the angular momentum operator. The coupling constant is given by $U=4\sqrt{\pi \lambda} N a/a_{\perp}$ with the particle number $N$, the s-wave scattering length $a$, and the aspect ratio $\lambda=\omega_{z}/\omega_{\perp}$ of the trapping potential. The pinning effect is caused by the OL potential $V_{\rm OL} = V_{0} [\sin^{2} (kX) + \sin^{2} (kY) ]$ with square symmetry. We consider the situation in which the OL rotates with a rotation frequency $\omega$ different from $\Omega$ of the condensate. Then, $V_{\rm OL}$ has an explicit time dependence through
\begin{eqnarray}
\left(
\begin{array}{c}
X \\
Y 
\end{array}
\right) =\left( 
\begin{array}{cc}
\cos \delta \omega t & \sin \delta \omega t \\
- \sin \delta \omega t & \cos \delta \omega t 
\end{array}
\right) \left(
\begin{array}{c} 
x \\ 
y
\end{array} 
\right).
\end{eqnarray} 
Here, $\delta \omega = \omega - \Omega$ is the relative rotation frequency between the condensate and the OL; the OL rotates faster (slower) than the condensate for $\delta \omega > 0$ ($\delta \omega < 0$). Finally, we introduce the phenomenological parameter $\gamma$ in the left side of Eq. (\ref{tdGPeq}), which represents the damping of the condensate motion caused by the noncondensed atoms rotating with $\Omega$ \cite{Tsubota}. The damping is necessary to obtain the phase diagram of the quasi-stationary states. We use the small value $\gamma = 0.01$ which is small enough to ensure the underdamped motion of the vortices; the overdamped motion is unlikely to occur in an ultracold dilute atomic BEC. 

\begin{figure}
\includegraphics[height=0.34\textheight]{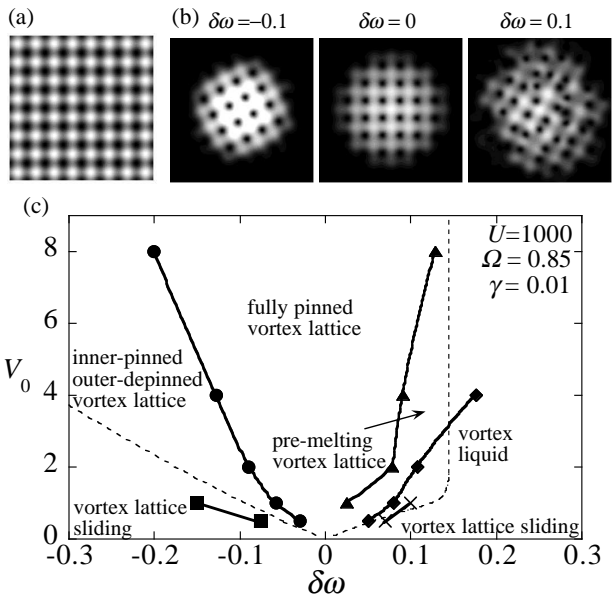}
\caption{(a) The profile of the OL for $k=\pi/3$ at $t=0$. Bright (Dark) regions represent the potential maxima (minima). (b) Condensate density for each $\delta \omega$ at a certain time in the quasi-stationary stage. The parameter values are $U=1000$ and $\Omega=0.85$. The middle panel ($\delta \omega = 0$) is the initial state of the simulations. The presented box dimensions along $x$ and $y$ are $-$11.8 to $+$11.8 (The simulations were done in the region $-$14.0 to $+$14.0).  (c) $\delta \omega$ (relative rotation frequency) - $V_{0}$ (potential amplitude) phase diagram of the dynamical vortex states. The phase boundary is determined by calculating the structure factor and the angular momentum per atom, as described in the text. The dashed boundary for the depinning of all vortices is obtained by solving Eq. (\ref{crideltame}).}
\label{phasedia}
\end{figure}
Our simulations start from the initial state in which the BEC is equilibrated under rotation $\Omega$ and a corotating OL with the same rotation frequency $\omega = \Omega$. Although a rich variety of equilibrium vortex structures emerge depending on the system parameters \cite{Pu}, we focus on the initial state in which vortices are {\it commensurate} with the pinning sites. For example, given $\Omega=0.85$, the wave number $k=\pi/3$ for the OL [see Fig. \ref{phasedia}(a)] and the coupling constant $U=1000$, we can obtain such an initial state through the imaginary time propagation of Eq. (\ref{tdGPeq}), as shown in the middle of Fig. \ref{phasedia}(b). The nucleated vortices are pinned at the potential maxima of the OL and form a square lattice. We do not consider the very small values of $V_{0}$ either, because the competition between triangular vortex lattice and the square OL potential occurs \cite{Tung}. Next, we tune the rotation frequency of the OL by $\delta \omega$ adiabatically so as not to excite the collective motion \cite{tyuu1}. We then calculate the time evolution from this initial state for a given set of parameters ($\delta \omega$, $V_{0}$), using the Crank-Nicolson implicit scheme. 

\begin{figure}
\includegraphics[height=0.295\textheight]{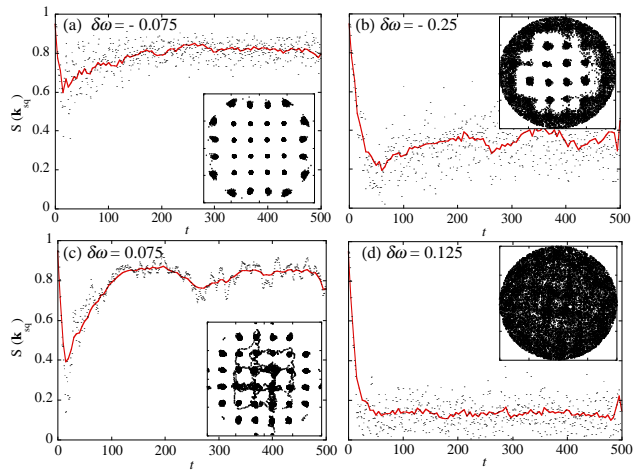}
\caption{(Color online) Time development of the structure factor $S({\bf k}_{\rm SQ}, t)$ for $V_{0}$=2.0 and for different values of $\delta \omega$. The average of the obtained data is given by the solid curve. The inset show the corresponding vortex trajectories within the Thomas-Fermi radius of the initial state in a frame rotating with the OL.}
\label{trajectry}
\end{figure}
\begin{figure}
\includegraphics[height=0.16\textheight]{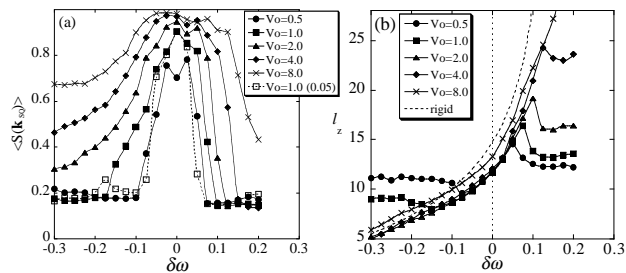}
\caption{Time-averaged values of the structure factor $\langle S({\bf k}_{\rm SQ}) \rangle$ (a) and the angular momentum per atom $\langle l_{z} \rangle$ (b) as a function of $\delta \omega$ for several values of $V_{0}$. In (a), the result for $\gamma=0.05$ and $V_{0}=1.0$ is also shown. In (b), the plot of the rigid-body estimation \cite{tyuu2} $l_{z}=C(\Omega+\delta \omega)/\sqrt{1-(\Omega+\delta \omega)}$ is shown by the dotted curve; $C=6.7$ is the fitting parameter.}
\label{angle}
\end{figure}

After a sufficiently long time, the condensate motion becomes quasi-stationary because of the damping parameter $\gamma$, with the remaining durable vortex dynamics driven by the OL. These dynamics exhibit nontrivial behavior depending on the values of ($\delta \omega$, $V_{0}$), summarized in the $\delta \omega$ - $V_{0}$ phase diagram in Fig. \ref{phasedia}(c). To characterize clearly the dynamics and the underlying structures, we calculate the vortex trajectory {\it in the frame rotating with the OL} and the corresponding structure factor defined as $S ({\bf k},t)= \frac{1}{N_{c}} \sum_{j} \exp [i {\bf k} \cdot {\bf r}_{j}(t)]$ \cite{Pu}, where $j$ labels individual vortices, ${\bf r}_{j}(t)$ is the position of the $j$th vortex and $N_{c}$ is the total number of vortices. If vortices are pinned by the OL and form a perfect square lattice, $S({\bf k},t)$ becomes unity if we choose ${\bf k} = {\bf k}_{\rm SQ} = 2 k \hat{\bf y}$, which is one of the two fundamental reciprocal lattices for the square lattice. Figure \ref{trajectry} shows the time evolution of $S ({\bf k}_{\rm SQ},t)$ as well as the vortex trajectories. Here, we pick out the vortices inside the Thomas-Fermi radius of the initial state to obtain Fig. \ref{trajectry}. The time-averaged value of $S({\bf k}_{\rm SQ},t)$ and the angular momentum per atom $l_{z}$ are the useful quantities to characterize each distinct phase, which are shown in Fig. \ref{angle}. 

There is a remarkable difference in the dynamics between the positive side of $\delta \omega$ and the negative one. For $\delta \omega < 0$, the rotation of the OL is slower than $\Omega$. If $|\delta \omega|$ is small, the initially pinned vortices stay in their original pinning sites, being dragged by the rotating OL. This behavior is shown in the inset of Fig. \ref{trajectry}(a); $S({\bf k}_{\rm SQ})$ stays about 0.8 in the stationary stage. Then, the condensate radius and the angular momentum of the condensate decreases [see Fig. \ref{phasedia}(b) and Fig. \ref{angle}(b)] because its rotation must be reduced by the dragged vortices. An increase in $|\delta \omega|$ causes depinning of vortices from the condensate periphery as shown in Fig. \ref{trajectry}(b), which leads to the decrease of the time-averaged value $\langle S({\bf k}_{\rm SQ}) \rangle$. With further increase in $|\delta \omega|$, the vortices cannot catch up to the motion of the OL and start to make a sliding motion. The onset of the sliding is seen from the deviation of $\langle l_{z} \rangle$ from the rigid-body estimation. When the sliding motion occurs, all vortices rest in the rotating frame of $\Omega$. We identify the vortex phase as follows: (i) ``fully pinned vortex lattice" which is determined by $\langle S({\bf k}_{\rm SQ}) \rangle > 0.7$, (ii) ``inner-pinned outer-depinned vortex lattice" by $0.25 < \langle S({\bf k}_{\rm SQ}) \rangle < 0.7$ and by $\langle l_{z} \rangle$ following the rigid-body estimation, and (iii) ``vortex lattice sliding" by $\langle S({\bf k}_{\rm SQ}) \rangle < 0.25$ and by $\langle l_{z} \rangle$ deviating from the rigid-body estimation. 

For $\delta \omega > 0$, if vortices are pinned, the angular momentum of the condensate must increase because the OL rotates faster than $\Omega$. This demands additional vortex nucleation to ensure the fast rotation. The nucleated vortices give the incommensurate effect for the vortex lattice. These vortices wander interstitially, exchanging their locations with the pinned vortices, where the vortex trajectories show that the vortices move to nearest-neighbor sites. For small $\delta \omega$, the vortex configuration keeps the original square symmetry, $\langle S({\bf k}_{\rm SQ})\rangle$ staying about 0.85. However, a slight increase in $\delta \omega$ rapidly decreases $\langle S({\bf k}_{\rm SQ})\rangle$, which indicates depinning of some or all commensurate vortices. We find that the interstitial vortices induce melting of the vortex lattice from the inside as shown in Fig. \ref{trajectry}(c). If $V_{0}$ is relatively small, $\langle l_{z} \rangle$ in the quasi-stationary stage is near the initial value and there are few interstitial vortices. The vortices then make a sliding motion. For large $V_{0}$ on the other hand, the stationary value of $\langle l_{z} \rangle$ is larger than the initial value and the incommensurate effect becomes important. Then, as shown in Fig. \ref{trajectry}(d), the vortex motion is fully disordered by the nucleated additional vortices which prevent the other vortices from pinning, exhibiting a liquid-like behavior. We characterize the phase (i) ``fully pinned vortex lattice" by $\langle S({\bf k}_{\rm SQ}) \rangle > 0.85$, (iv) ``pre-melting vortex lattice" by $0.25< \langle S({\bf k}_{\rm SQ}) \rangle < 0.85$, (v) ``vortex liquid" by $\langle S({\bf k}_{\rm SQ}) \rangle < 0.25$ and $\langle l_{z} \rangle-l_{z}(t=0)>2$, and (vi) ``vortex lattice sliding" by $\langle S({\bf k}_{\rm SQ}) \rangle < 0.25$ and $\langle l_{z} \rangle-l_{z}(t=0)<2$. 

The onset of these different phases can be described using force balance arguments. The force acting on a vortex is given by ${\bf F}={\bf F}_{d}+{\bf F}_{p}+{\bf F}_{vv}$, where ${\bf F}_{d}$ is the driving force, ${\bf F}_{p}$ the pinning force, and ${\bf F}_{vv}$ the force from the other vortices. The commensurate vortex will remain pinned as long as the inequality $|{\bf F}_{p}| >| {\bf F}_{d}|  + |{\bf F}_{vv}|$ is satisfied. The pinning potential $U_{\rm pin}$ for the vortex displacement ${\bf r}=(x, y)$ was calculated by Reijnders and Duine \cite{Reijnders} for the case without a harmonic trap. Using their results, we obtain the magnitude of the pinning force ${\bf F}_{p} = - \nabla U_{\rm pin}$ as 
\begin{equation}
|{\bf F}_{p}| = \frac{a_{z}}{4a} k V_{0} Q(k\xi) \sqrt{\sin^{2}(2 k x) + \sin^{2}(2 k y) },
\end{equation}
where $a_{z}$ is the size along the $z$-axis, $\xi =( 8 \pi a \rho )^{-1/2}$ the healing length with the condensate density $\rho$, and $Q(k\xi)=\int^{\infty}_{\xi} dr J_{0}(2 k r) / r + J_{1}(2 k \xi) / 2 k \xi$ with the $l$th-order Bessel function $J_{l}$ of the first kind. This analytic form is valid when the vortex core is much smaller than an optical lattice period and the strength of the potential is sufficiently weak. 

For $\delta \omega<0$, because the incommensurate effect is not important, the square symmetry of the pinning site makes the ${\bf F}_{vv}$ irrelevant for the vortex dynamics. The driving force is given by the sum of the Magnus force ${\bf F}_{M}$ and the mutual friction force ${\bf F}_{D}$; the mutual friction originates from the interaction between the vortex core and normal-fluid flow. If a vortex is pinned by the rotating OL, the vortex experiences the background flow with $- \delta \omega \hat{\bf \theta}$ {\it in the rotating frame of $\omega$}. Then, the Magnus force is written as ${\bf F}_{M} = a_{z} m \rho \kappa \delta \omega r \hat{\bf z} \times \hat{\bf \theta}$ with the quantum circulation $\kappa=h/m$. Since the Magnus force becomes larger as $r$ increases, vortices begin to depin from the condensate periphery. At near zero temperature, the thermal atoms and the resulting mutual friction force can be negligible. Hence, the sliding motion can occur when $|{\bf F}_{P}|$ for vortices near the trap center is less than $|{\bf F}_{D}|$ on the same position. We focus on a vortex located at the pinning site $(x,y) = (\pi/2k,\pi/2k)$. By comparing the ${\bf F}_{M}$ at $r=\pi/\sqrt{2} k$ and the maximum of $|{\bf F}_{p}|$ at $(x,y) = (\pi/4k,\pi/4k)$, we can obtain the critical $\delta \omega$ for the depinning of all vortices. To estimate the chemical potential $\mu$ and the healing length $\xi$, we use the centrifugal-force modified Thomas-Fermi approximation that neglect the density variation caused by the vortex cores. Using the central density $\rho = m \mu/ 4 \pi \hbar^{2} a$, the critical $\delta \omega$ can be obtained by solving the equation
\begin{equation}
\frac{2}{\pi} k^{2} V_{0} Q(k\xi(\delta \omega)) = \sqrt{\frac{U (1-\delta \omega - \Omega)}{2 \pi}} \delta \omega. \label{crideltame}
\end{equation}
The result is shown by the dashed line in Fig. \ref{phasedia}(c), which is in reasonable agreement with the numerically obtained boundary for the sliding motion. 

For $\delta \omega > 0$, the argument becomes more complex because the interactions caused by the interstitial vortices cannot be negligible; the solution of Eq. (\ref{crideltame}) overestimate the boundary of the vortex-liquid phase because of the neglect of ${\bf F}_{vv}$. Also, the Thomas-Fermi approximation breaks down for $\Omega + \delta \omega = \omega \simeq 1$ ($\delta \omega \simeq 0.15$). We need further detailed consideration to obtain the onset of the vortex lattice melting caused by the incommensurate effect. 

The melting of vortex lattices is one of the most important topics in atomic-gas BECs. The melting was predicted to occur in the limit of fast rotation and low condensate particle number through thermal \cite{Gifford} or quantum fluctuations \cite{Sinova}. The melting mechanism shown in this paper is different from those studies; the vortex liquid phase is supported by the driving force and the dissipation. This is one of the examples of dissipative structures sustained by the balance between energy input and output \cite{Nicolis}. 

Using different values of the dissipation parameter $\gamma$,  we find no qualitative change in the results if the value of $\gamma$ is sufficiently small. However, with increasing $\gamma$, we find that the region of the fully-pinned lattice is reduced from the $\gamma=0.01$ results; an example of $\langle S({\bf k}_{\rm SQ}) \rangle$ with $\gamma=0.05$ and $V_{0}=1.0$ is shown in Fig. \ref{angle}(a). This is due to the increase of the driving force associated with the mutual friction force ${\bf F}_{D}$; the damping parameter $\gamma$ is related to the interaction between the condensate atoms and the thermal atoms \cite{Tsubota}. 

In conclusion, we study the dynamical vortex phases of a trapped BEC in the presence of a rotating OL with a different rotation frequency from the condensate. We find several nonequilibrium dissipative structures, where the vortices exhibit complex dynamics due to the competition of the driving force, the pinning force and the vortex-vortex interaction. When the rotation rate of the OL becomes higher than that of the condensate, the additional nucleated vortices destroy the commensurate vortex lattice, leading to the vortex liquid-like phase.  

The authors are grateful to P. Louis for a critical reading of this manuscript. K.K. acknowledges the supports of Grant-in-Aid for Scientific Research from JSPS (Grant No. 18740213). M.T. acknowledges the supports of Grant-in-Aid for Scientific Research from JSPS (Grant No. 18340109) and Grant-in-Aid for Scientific Research on Priority Areas (Grant No. 17071008) from MEXT.



\begin{thebibliography}{99}
\bibitem{Braun}
O. M. Braun and Yu. S. Kivshar, {\it The Frenkel-Kontorova Model}, Springer, Berlin (2004).
\bibitem{Reijnders}
J.W. Reijnders and R.A. Duine, Phys. Rev. Lett. {\bf 93}, 060401 (2004); Phys. Rev. A {\bf 71}, 063607 (2005). 
\bibitem{Pu}
H. Pu, L.O. Baksmaty, S. Yi, and N.P. Bigelow, Phys. Rev. Lett. {\bf 94}, 190401 (2005). 
\bibitem{Wu}
C. Wu, H. Chou, J. Hu, and S.C. Zhang, Phys. Rev. A {\bf 69}, 043609 (2004).
\bibitem{Bhat}
R. Bhat, L.D. Carr, and M.J. Holland, Phys. Rev. Lett. {\bf 96}, 060405 (2006).
\bibitem{Tung}
S. Tung, V. Schweikhard, and E.A. Cornell, cond-mat/0607697.
\bibitem{Madison}
K.W. Madison, F. Chevy, W. Wohlleben, and J. Dalibard, Phys. Rev. Lett. {\bf 84}, 806 (2000);
J.R. Abo-Shaeer, C. Raman, J.M. Vogels, and W. Ketterle, Science, {\bf 292}, 476 (2001);
I. Coddington, P. Engels, V. Schweikhard, and E.A. Cornell, Phys. Rev. Lett. {\bf 91}, 100402 (2003);
\bibitem{Morsch}
O. Morsch and M. Oberthaler, Rev. Mod. Phys. {\bf 78}, 179 (2006).
\bibitem{Nicolis}
G. Nicolis and I. Prigogine, {\it Self-Organization in Non-Equilibrium Systems}, Wiley, New York (1977).
\bibitem{Pethickbook}
C.J. Pethick and H. Smith, {\it Bose-Einstein Condensation in Dilute Gases}, Cambridge University Press, Cambridge (2002).
\bibitem{Tsubota}
K. Kasamatsu, M. Tsubota, and M. Ueda, Phys. Rev. A {\bf 67}, 033610 (2003). 
\bibitem{tyuu1}
The relative rotation frequency is ramped up as $\delta  \omega = 0.005 t \Theta(\delta \omega_{f} - \delta \omega)$, where $\Theta(x)$ is a step function. Throughout the paper, the value of $\delta \omega$ represents the final value $\delta \omega$.
\bibitem{tyuu2}
For rigid-body rotation, the angular momentum per atom is given by $l_{z} = L_{z} /\hbar N = (1/2) \int d{\bf r} (\Omega-\delta \omega) r^{2} |\psi|^{2}$ in our dimensionless scale. Using the centrifugal-force modified Thomas-Fermi radius for the integration region and neglecting the spatial variation of the density due to the vortex cores and the OL, we obtain $l_{z} = (\sqrt{2 U/\pi}/3) (\Omega - \delta \omega)/\sqrt{1-(\Omega -\delta \omega)}$. This result differs from the numerical results by a few factor. 
\bibitem{Gifford}
S.A. Gifford and G. Baym, Phys. Rev. A {\bf 70}, 033602 (2004). 
\bibitem{Sinova}
N.R. Cooper, N.K. Wilkin, and J.M.F. Gunn, Phys. Rev. Lett. {\bf 87}, 120405 (2001); J. Sinova, C.B. Hanna, and A.H. MacDonald, {\it ibid.} {\bf 89}, 030403 (2002); N. Regnault and Th. Jolicoeur, {\it ibid.} {\bf 91}, 030402 (2003); M. Snoek and H.T.C. Stoof, {\it ibid.} {\bf 96}, 230402 (2006).
\end{thebibliography}
\end{document}